\documentclass[]{aa} % \usepackage{graphics}
\usepackage{psfig}

\newcommand{\Zs}{$Z_{\odot}$}
\newcommand{\Ms}{$M_{\odot}$}
\newcommand{\mstar}{$M_{\star}$}

\newcommand{\msun}{\ifmmode M_{\odot} \else M$_{\odot}$\fi}
\newcommand{\rsun}{\ifmmode R_{\odot} \else R$_{\odot}$\fi}
\newcommand{\lsun}{\ifmmode L_{\odot} \else L$_{\odot}$\fi}
\newcommand{\zsun}{\ifmmode Z_{\odot} \else $Z_{\odot}$\fi}

\newcommand{\qh}{$Q({\rm{H^{0}}})$}

\newcommand{\Toiii}{$T_{{\rm [O\ III]}4363/5007}$}

\newcommand{\Mup}{$M_{\rm up}$}

\newcommand{\Ha}{H$\alpha$}
\newcommand{\Hb}{\ifmmode {\rm H}\beta
\else H$\beta$\fi}

\newcommand{\hii}{H\ {\sc ii}}

\newcommand{\Oii}{[O\ {\sc ii}]$\lambda$3727}
\newcommand{\oii}{[O\ {\sc ii}]}
\newcommand{\Oiii}{[O\ {\sc iii}]$\lambda$5007}
\newcommand{\Oiiit}{[O\ {\sc iii}]$\lambda$4363}
\newcommand{\oiii}{[O\ {\sc iii}]}

\newcommand{\rSii}{[S\ {\sc ii}]$\lambda$6731/6717}

\newcommand{\Hp}{H$^{+}$}

  \newcommand{\Hep}{He$^{+}$}
\newcommand{\Hepp}{He$^{++}$}

\begin{document}

%    \thesaurus{11 % A&A Section 11: Galaxies
%               (11.01.1; %Galaxies: abundances,
%                11.09.1; %Galaxies: individual
%                11.09.4; % Galaxies: ISM,
%                11.19.3; %Galaxies: starburst,
%                08.05.1; %Stars: early-type,
%                08.23.2)} %Stars: Wolf-Rayet,

%
     \title{Collisional excitation of hydrogen and the determination of
the primordial helium abundance from \hii\ regions}

     \author{Gra\.{z}yna Stasi\'{n}ska \inst{1}, Yuri Izotov \inst{2}}

\institute{ DAEC, Observatoire de Paris-Meudon, F-92195 Meudon Cedex,
   France\\email: grazyna.stasinska@obspm.fr\and Main Astronomical Observatory,
Ukrainian National Academy of Sciences, Kyiv 03680,  Ukraine \\
email: izotov@mao.kiev.ua} 
\date{*** / ***} 
\titlerunning{Primordial helium abundance determination}
\authorrunning{Stasi\'{n}ska \& Izotov} \offprints{G.\ Stasi\'{n}ska}

\abstract{
This paper investigates the effect of 
collisional enhancement of the hydrogen lines on the 
derivation of the helium abundances in low metallicity \hii\ regions. 
For this, we have constructed a grid of photoionization models relevant 
for the analysis of giant \hii\ regions in blue compact galaxies.
We show that the effect of collisional excitation on the  \Ha/\Hb\ ratio 
can be quite important (up to 
 8\% or more). The impact of this effect on the determination of 
the helium mass fraction has been tracked on four 
low-metallicity blue compact galaxies for which Keck spectra 
are available and which are among the best objects for the 
quest of the pregalactic helium abundance.  
We find that taking into account the effects of collisional excitation of hydrogen  
results in an upward correction of the helium mass fraction $Y$ by
up to  5\%.
However, combining with other systematic effects usually not considered in the
determination of the helium abundance in low-metallicity
galaxies, 
the resulting uncertainty should be much less. 
%***keywords have to be revised***
 }

\maketitle

\section{Introduction}

Tests of the big bang models and the determination of cosmological
parameters require an accurate determination of the relative abundances
of the light isotopes produced during the big bang (Steigman et al. 1977; 
Walker et al. 1991; Sarkar 1996).
The abundance of primordial $^4$He is mostly derived from the
helium abundances in metal-poor extragalactic
   \hii\ regions (Kunth \& Sargent 1983; Pagel et al. 1992; Izotov \& Thuan
1998) although other nebulae such as
   \hii\ regions in the Magellanic Clouds (Peimbert et al. 2000) or
planetary nebulae in the Galactic halo (Peimbert 1983, 1989; Clegg 1989)
have also occasionally been used. This is a challenging task, since
the required accuracy is on the one percent level.

In order to achieve such a goal, excellent quality observations are
obviously required together with an extremely reliable reduction
procedure.  With the advent of very large telescopes and the progress
in detectors, recent years have produced a large amount of data
suitable for such an enterprise (Izotov et al. 1997; Izotov
\& Thuan 1998; Izotov et al. 1999).  As regards the
interpretation of the observed line intensities in terms of abundance
ratios, many problems occur and still remain to be overcome.  For
example, the atomic data relevant for the helium line emissivities
have only recently reached a state compatible with the demand (Benjamin
et al. 1999). Among the problems to be solved for the determination of
the helium abundance in one nebula, and towards which various groups of
researchers have had different approaches, are: the stellar absorption
lines underlying the nebular emission lines, the reddening correction,
the distribution of the electron temperature inside the nebulae, the
density structure and its effect on the line emissivities, the optical
depth in the lines, the ionization correction to derive
the He/H ratio from the measured \Hep/\Hp\ ratio. The next step is to
derive the pregalactic helium abundance from the measurement of the
helium abundance in a sample of galaxies, by extrapolating the
results towards zero metallicity. These problems have been
extensively discussed in a long list of papers including
Peimbert \& Torres-Peimbert (1974, 1976), Stasi\'nska (1980), Rayo et al. (1982),
 Kunth \& Sargent
(1983), Davidson \& Kinman (1985), Dinerstein \& Shields (1986), Pagel et al.
(1986, 1992), Campbell (1992), Skillman \& Kennicutt (1993), Balbes et al.
(1993), Mathews et al. (1993), Olive \& Steigman (1995), Sasselov \&
Goldwirth (1995), Olive et al. (1997), Steigman et al. (1997), Izotov \&
Thuan (1998), Izotov et al. (1997, 1999), Armour et al. (1999), 
Ballantyne et al. (2000), Viegas et al. (2000), Peimbert et al. (2000, 2001), 
Olive \& Skillman (2001), Sauer \& Jedamzik (2001).

The present day situation is that different groups reach different
estimates of the pregalactic helium abundance, mutually exclusive
within the quoted error bars.  For example, Olive et al. (1997) give
$Y_{\rm p}$ = 0.234$\pm$0.002, while Izotov \& Thuan (1998) give 
$Y_{\rm p}$ = 0.244$\pm$0.002.  These two estimates lead to
very different cosmological implications, as discussed by Izotov et al.
(1999).

The effect on the pregalactic helium abundance determination arising
from deviation from case B theory for the hydrogen lines has been
relatively little discussed so far. The effect of finite optical depth
and dust on \Hb\ emissivity have been discussed by Cota \& Ferland
(1988) and Hummer \& Storey (1992). 
The effect of collisional excitation of the hydrogen lines by
thermal electrons has been pointed out by Davidson \& Kinman (1985)
and discussed for the case of I Zw 18 by Sasselov \& Goldwirth (1995)
and Stasi\'nska \& Schaerer (1999). Stasi\'nska (2001) 
has shown that this
effect can lead to an underestimation of the He {\sc i} 5876/\Hb\ ratio 
by as much as  10\% in hot planetary nebulae, independent of the actual reddening of the
object.

   The purpose of the present paper is to systematically explore the
   effects of collisional excitation of the hydrogen lines on the helium
   abundance determinations in metal-poor giant \hii\ regions. For this,
   we use a grid of photoionization models of giant \hii\ regions and
   consider them in the light of
   the spectral properties of a sample of blue compact \hii\ galaxies.
This allows us to obtain a realistic estimate of
  the effect of collisional excitation by thermal
electrons both on \Ha\ and on \Hb.  We then concentrate on a selection of
blue compact galaxies with particularly good quality spectroscopic
data, and track the effect of collisional excitation - at the level
predicted by the models - on the derived helium abundance for these
objects.

\section{The photoionization models}

The photoionization models of giant \hii\ regions are constructed
using the procedure described in  Stasi\'nska et al. (2001, hereinafter SSL2001)
 to which we refer for details.
Briefly, we consider coeval stellar clusters of given initial
mass \mstar\ that ionize the surrounding gas, supposed to be of
same chemical composition as the stars. We assume
 instantaneous bursts with a Salpeter
initial mass function and an upper stellar limit \Mup\ = 120 \Ms.
The radiation field is provided by the evolutionary synthesis
models of Schaerer \& Vacca (1998)
based on the non-rotating Geneva stellar evolution models, with the
high mass-loss tracks of Meynet et al.\ (1994).  The spectral energy
distributions for massive main-sequence stars are those given by the
{\em CoStar} models (Schaerer \& de Koter 1997).  The pure
He models of Schmutz et al.\ (1992) are used for Wolf-Rayet stars.
The ionization and temperature structure of the nebula is computed
with the photoionization code
PHOTO, as described in Stasi\'nska \& Leitherer (1996, hereinafter
SL96). We assume that the nebulae are spherical and radiation-bounded, 
with uniform density $n$ =
10~cm$^{-3}$ and filling factor $\epsilon$ = 1 
and that all the stars are located at the center.
As in our previous papers (SL96 and SSL2001) and in all the papers
based on static photoionization models, our models
do not consider the possible pollution of the nebulae 
by stellar ejecta during the course of the evolution of the star clusters.
 While this phenomenon may have important consequences, 
it is not really relevant to the problem under study here.

%**********************************
%     Fig1
%**********************************
\begin{figure*}[!ht]
\centerline{\psfig{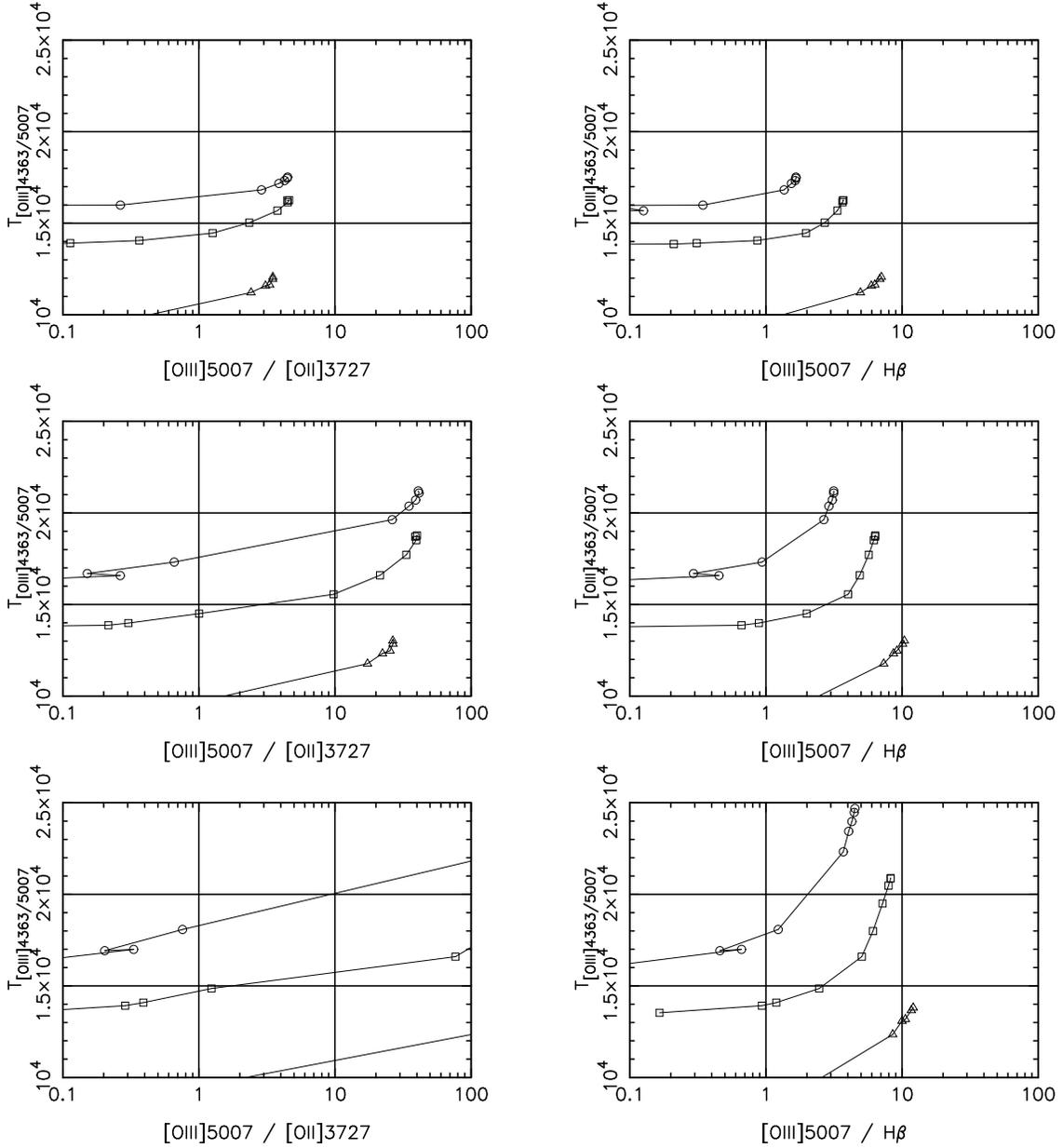}}
\caption{Sequences of photoionization models for
instantaneous bursts of various metallicities. 
The metallicity is indicated by the following symbols: circle for $Z$/\Zs\ =
0.02, square for $Z$/\Zs\ = 0.05, triangle for $Z$/\Zs\ = 0.2.
The symbols mark time
steps of 1 Myr. 
The models shown in the top panels have \mstar = 10$^3$ \Ms, 
those in the middle panels have \mstar = 10$^6$ \Ms, 
those in the bottom panels have \mstar = 10$^9$ \Ms\ (see text).
The ordinate represents the temperature indicated by the \Oiiit/\Oiii\ line ratio.
In the left panels it is shown as a function of \Oiii/\Oii, 
in the right panels as a function of \Oiii/\Hb.\label{Fig1}}
\end{figure*}
%************************************

The grid used in the present study is composed of models with
metallicities $Z$ of 0.2, 0.05 and 0.02 times the solar metallicity
(using the prescription of McGaugh (1991) for the relative abundances
of the elements). 
As is known (e.g. Shields 1986) the emission line spectrum of photoionized nebulae 
is essentially dependent on the spectral energy distribution 
of the ionizing radiation field, on the metallicity and on the 
ionization parameter $U$. For a constant density sphere $U$
is equal to $A (Q({\rm{H^{0}}}) n \epsilon^{2})^{1/3}$, 
where \qh\ is the total number
of H Lyman continuum photons emitted by the ionizing source 
(proportional to the mass of the star cluster) and
$A$ is approximately given by 
$A \approx 2.8 \times 10^{-20} {(10^{4}/T_{\rm e})}^{2/3}$, 
where $T_{\rm e}$ is the electron temperature.
% $A$ depends only on the electron temperature (see e.g. SL96). 
At a given chemical composition, any combination of 
 \qh, $n$ and $\epsilon$ giving the same value of $U$ will 
result in the same emission line spectrum (at least when the 
densities are sufficiently
low that cooling is not affected by collisional deexcitation of forbidden lines, 
which is the case of the objects under study in this paper).
 Three
different initial masses for the star
clusters are considered:
  \mstar=10$^{3}$, 10$^{6}$,
and 10$^{9}$\Ms\ (assuming a lower IMF mass cut-off of $M_{\rm low}$ =
   0.8\Ms). With such a grid, we reasonably cover the range 
of metallicities and ionization parameters relevant to the objects we are interested in.

For each model, we derive the electron temperature \Toiii\ 
from the computed \Oiiit/\Oiii\
ratio\footnote{
Throughout the paper, [O{\sc iii}]$\lambda$5007  refers to the intensity of the line at 5007\AA, 
not to the sum of the lines
 [O {\sc iii}]$\lambda$4959 and $\lambda$5007.}, 
using the same atomic data as in the photoionization code.
Figure 1 presents the sequences of models in the \Toiii\  versus
\Oiii/\Oii\ plane (left) and in the \Toiii\ versus
\Oiii/\Hb\ plane (right). The top panels correspond to a total initial
stellar mass of \mstar=10$^{3}$\Ms, the middle ones to \mstar=10$^{6}$\Ms\ 
and the bottom ones to \mstar=10$^{9}$\Ms. Symbols mark the epoch by time
steps of 10$^{6}$ yr,
starting from an age of 10$^{4}$ yr. Series with $Z$= 0.02
\zsun\ are represented by circles, series with $Z$= 0.05
\zsun\ by squares, and series with $Z$= 0.2
\zsun\ by triangles. The time evolution is from the right to the left in 
these plots, since the ionization parameter decreases due
to gradual disappearance of the most massive stars. This provokes a 
softening of the 
ionizing radiation field, which induces a decrease in the electron temperature.
 As expected, 
the highest temperatures are found in the  models with lowest metallicities 
(cooling by collisional excitation of forbidden lines becomes less important) 
and highest ionization parameters 
(cooling by collisional excitation of H Lyman $\alpha$ becomes less important).

In the next section, we discuss the relevance of this grid of models
for the problem we are investigating in the present paper.

\section{The sample of blue compact galaxies confronted to the 
model grid}

%****************************************
%   Fig2
%****************************************
\begin{figure*}
\centerline{\psfig{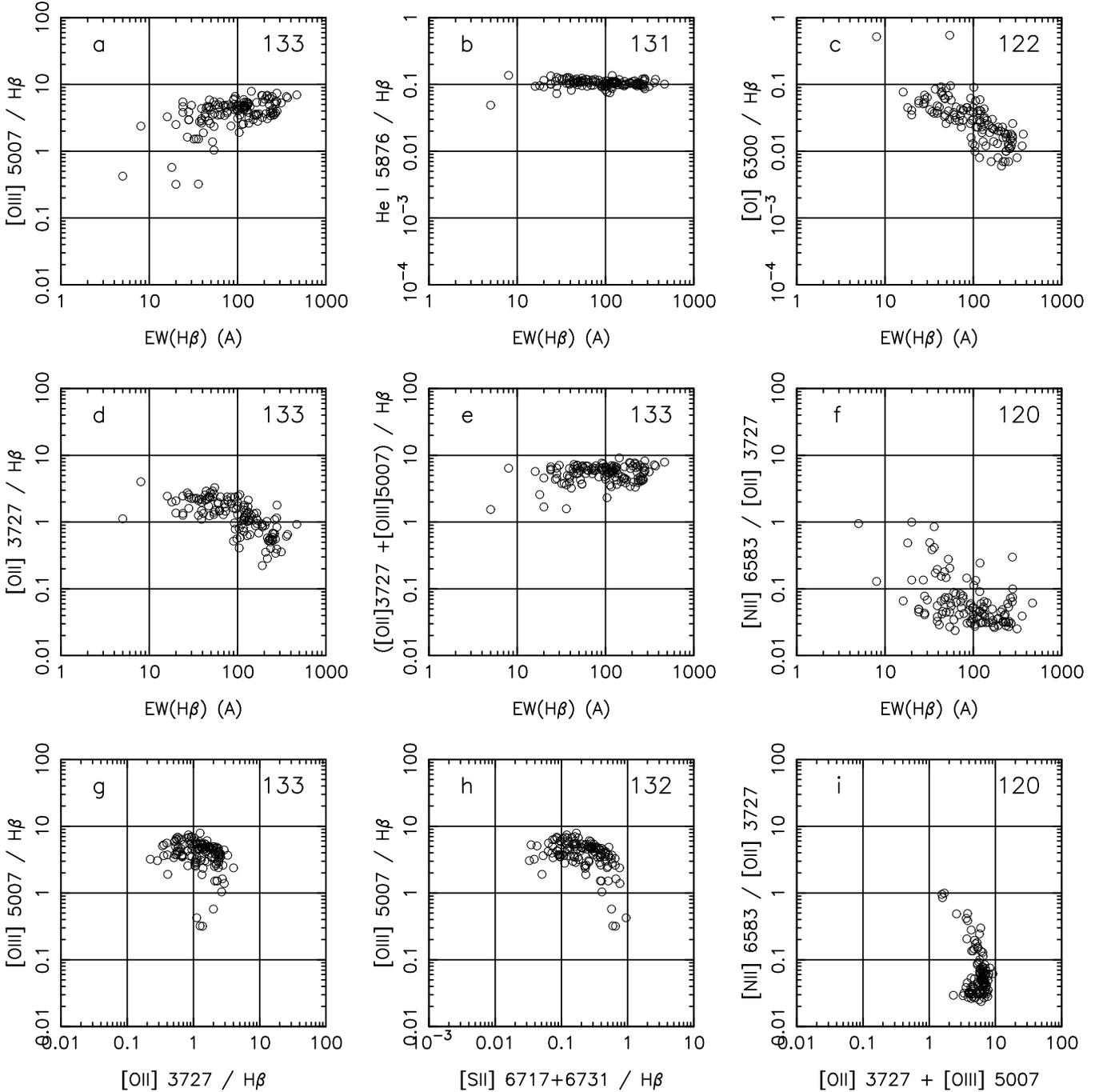}}
\caption{Our
observational sample in various diagrams relating emission line 
ratios and \Hb\ equivalent width. 
On the upper right of each panel is given the total number of the objects 
appearing in the diagram.
The behaviour of the present sample in these diagrams 
is very similar to the behaviour of the smaller sample shown by Stasi\'nska et al. (2001)
in their Fig. 2. \label{Fig2}}
\end{figure*}
%****************************************

Since ab initio models of nebulae photoionized by evolving starbursts
do not necessarily well
represent the observed properties of giant \hii\ regions
(see discussion in SSL2001), we have to use them in
the light of observed properties of \hii\ galaxies. For this, we consider 
a sample
of blue compact galaxies observed since 1994 by Izotov and coworkers.
In addition to the 69 objects considered in the study
by SSL2001, it contains 64 new objects observed with
different telescopes, including Keck, VLT, MMT, 4m and 2.1m KPNO
(Izotov et al. 2001a,b; Guseva et al. 2001; Papaderos et al. 2001). The new 
observations were reduced in the same way as those considered by SSL2001. In
all cases, the line intensities have been corrected for reddening by
using a procedure described in Izotov et al. (1994) in which the
reddening constant and stellar absorption in the Balmer lines are
determined simultaneously.
Figure 2 shows the same observational diagrams as Fig. 2 of
SSL2001 for the new sample of 133 objects.
Exactly the same trends are seen (with a slightly higher dispersion
due to the fact that some of the new data are of lower quality than
those of the sample considered by SSL2001). Therefore, we are dealing with
the same population of objects. Hence the conclusions of SSL2001
still apply: the sample is mostly composed of objects whose
ionization is produced by a starburst of age younger than about 5 Myr.
In the following, we will then restrict our considerations to models corresponding to
such ages.

%*********************************
%   Fig3
%*********************************
\begin{figure*}
\centerline{\psfig{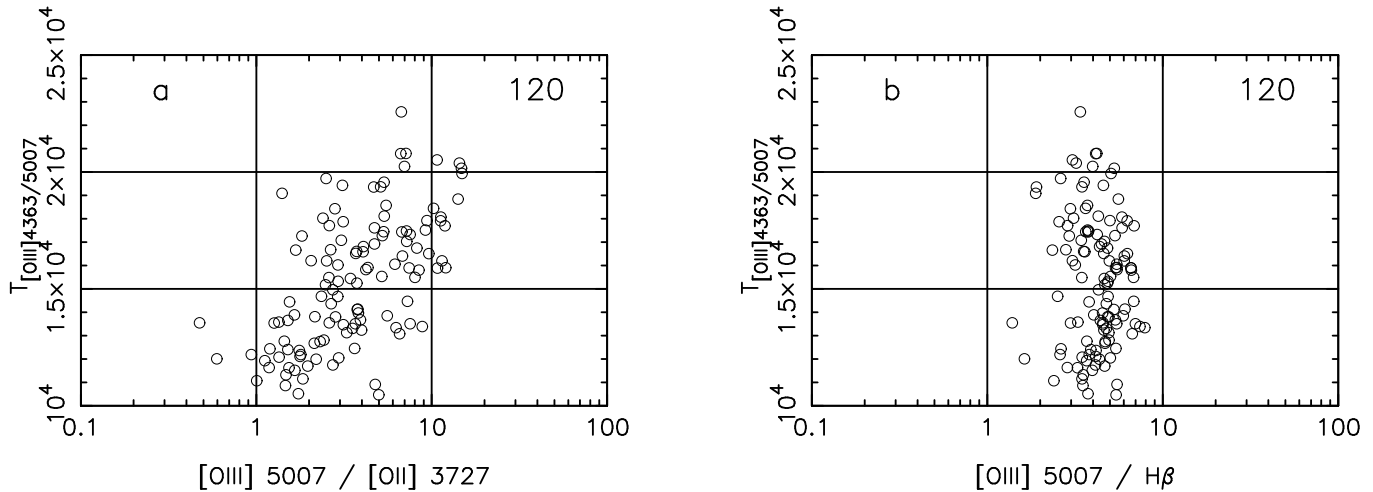}}
\caption{Our
observational sample in the same planes as the models shown in Fig. 1.
\label{Fig3}}
\end{figure*}
%*********************************

Figure 3 shows the observed sample in the \Toiii\  versus
\Oiii/\Oii\ plane (left) and in the \Toiii\  versus
\Oiii/\Hb\ plane (right). By comparing with Fig.
1, we see that  while the models with the highest \Toiii\  well reproduce
the \Oiii/\Hb\ ratios, they have 
\Oiii/\Oii\
ratios much larger than observed in objects of similar values of
\Toiii. For the range of observed values of \Oiii/\Oii, it is
impossible to obtain models with \Toiii\ around 20\,000K. 
One possible way to increase
the \oii\ line intensity is to consider composite models, in which
part of the emission comes from higher density filaments. However, 
it was impossible to find a prescription for the 
 composite models  to
reproduce at the same time \Oiiit/\Oiii, \Oiii/\Oii\ and the
electron density sensitive ratio \rSii. The conclusion is that simple
photoionization models are unable to reproduce the high
\Oiiit/\Oiii\ ratios that are observed in quite a number of giant \hii\ 
regions. A similar conclusion has been reached from
tailored model fitting of various giant \hii\ regions (Campbell 1988; 
Stasi\'nska \& Schaerer 1999; Luridiana
et al. 1999; Luridiana \& Peimbert 2001). Reducing the abundances of
coolants such as carbon or silicon (on which there are virtually no
observational constrains) is insufficient to solve the problem.  It
seems that an additional heating mechanism is at work, and a popular
candidate is shock heating. Whatever may be the origin of the
additional heating, this means that simple photoionization models
returning a \Oiii/\Oii\ in the observed range are likely to have too low
an electron temperature, and that those returning a \Oiiit/\Oiii\
comparable with observations have too high a degree of ionization.
This means that the amount of collisional excitation they predict for
the hydrogen lines is likely a lower limit, since the contribution of 
collisional
excitation increases with electron temperature and with the
proportion of residual neutral hydrogen.

\section{Model predictions for the collisional excitation of hydrogen lines}

%***************************
%    Fig4
%***************************
\begin{figure*}
\centerline{\psfig{figure=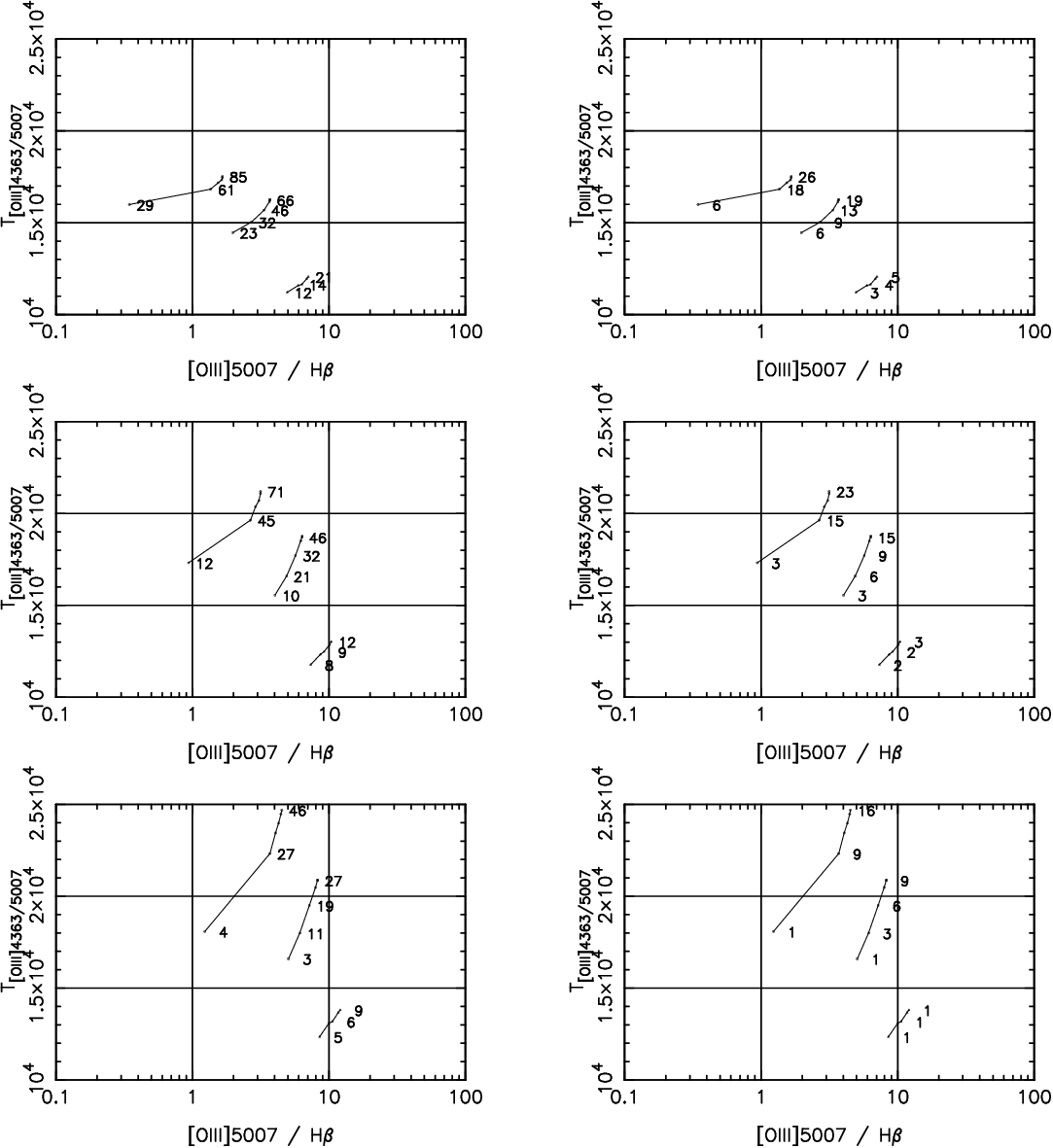,width=15.cm}}
\caption{The same models as in Fig. 1 shown in the 
\Toiii\ versus \Oiii/\Hb\ plane, but only for starburst ages up to 5 Myr. 
As in Fig. 1, the models shown in the top panels have \mstar = 10$^3$ \Ms, 
those in the middle panels have \mstar = 10$^6$ \Ms, 
those in the bottom panels have \mstar = 10$^9$ \Ms.
The numbers indicate, in units of 1 per 1000, the values of deviation from case B theory
for \Ha/\Hb\ (left panels) and \Hb\ (right panels). 
See text for a more detailed explanation. 
\label{Fig4}}
\end{figure*}
%****************************

In Figure 4, we display the same sequences of models as in Figure 1,
but up to ages of 5 Myr, in the  \Toiii\  versus
\Oiii/\Hb\  plane. Again, the top panels correspond to a total initial
stellar mass of \mstar=10$^{3}$\Ms, the middle ones to \mstar=10$^{6}$\Ms\ 
and the bottom ones to \mstar=10$^{9}$\Ms. The collisional contribution
to the Balmer line emission is indicated in the following way. In the
left panels, next to the models are indicated the values of :
\begin{displaymath}
1000\times \left(\frac{({\rm H}\alpha/{\rm H}\beta) {\rm (model)}}
{({\rm H}\alpha/{\rm H}\beta) {\rm (case\ B )}} - 1\right)
\end{displaymath}
%1000 $\times$ ((\Ha/\Hb) {(model) /  (\Ha/\Hb) (case B )) - 1)
where the case B value of \Ha/\Hb\ is computed for the temperature
\Toiii\ deduced from the \Oiiit/\Oiii\ ratio predicted by the model.

  It is seen that in our models the \Ha/\Hb\ ratios may be larger
  than the case B value (at the temperature \Toiii) by up to 8\%. This
  means that the reddening computed assuming the case B value of 
the \Ha/\Hb\ ratios may be significantly overestimated.

  In the right panels, next to the models, we indicate 
(in units of 1 per 1000) the collisional
  contribution to the emission in the \Hb\ line:  

\begin{displaymath}
1000\times\left(\frac{{\rm H}\beta {\rm (model)}}{{\rm H}\beta {\rm (case\ B)}}-1\right)
\nonumber
\end{displaymath}
%  $1000 \times (\Hb (model) /  \Hb (case B)) - 1)$
where \Hb (case B) stands for the \emph{integrated} emission of
the model in the \Hb\ line at the \emph{local} electron temperature.

Clearly, the collisional contribution to \Hb\ in our models may reach 2\%.

The computations in PHOTO have been made with the collisional
excitation rate coefficients from Aggarwal (1983) for the $n = 3$
level of hydrogen, and from Drake \& Ulrich (1980) for higher levels.
Values given by other authors can be significantly different. For
example Aggarwal et al. (1991) give values that are larger than the ones from
Drake \& Ulrich by up to a factor 2. Unfortunately, there is so far no
agreement on what are the most
reliable values of the collision strengths for hydrogen (Callaway 1994).

Note that there is no simple relation between the \emph{total} \Ha\  
and \Hb\ emission
in our models, although at each point in the nebula the relation
between the \Ha\ and \Hb\ emissivities is a function of the local
temperature only. The reason is that the electron temperature in our models
is not uniform, and the relation between local \Ha\ and \Hb\
emissivities do not propagate in a simple way into \emph{integrated} line
ratios. It is for a similar reason that a temperature correction factor 
\emph{tcf} has been introduced by Sauer \& Jedamzik (2001) to derive the helium 
abundance from the ratios of helium to hydrogen line intensities.

In real nebulae, where the temperature structure is
much more erratic than in simple photoionization models, 
the situation is probably even more complicated.

One might be tempted to use the ratio of H Lyman $\alpha$  to 
 H$\beta$ to better constrain the amount of collisional 
excitation of hydrogen lines in real nebulae. However,
 this proves difficult, because 
the path of H Lyman $\alpha$ photons is strongly increased by scattering in 
the ionized as well as in the neutral part of the nebula, during which absorption 
 by dust and molecular hydrogen occur. Therefore, the observed intensity of  
H Lyman $\alpha$
strongly depends on the gas and dust distribution and on the velocity field.

\section{The effect of H collisional excitation on the derived helium
abundances for a few selected blue compact galaxies}

It is instructive to investigate what are the
effects of collisional excitation of the hydrogen lines on the derived
helium abundance. This is done in the present section on a few examples.

%**************************************
%     Fig5
%**************************************
\begin{figure*}
\centerline{\psfig{figure=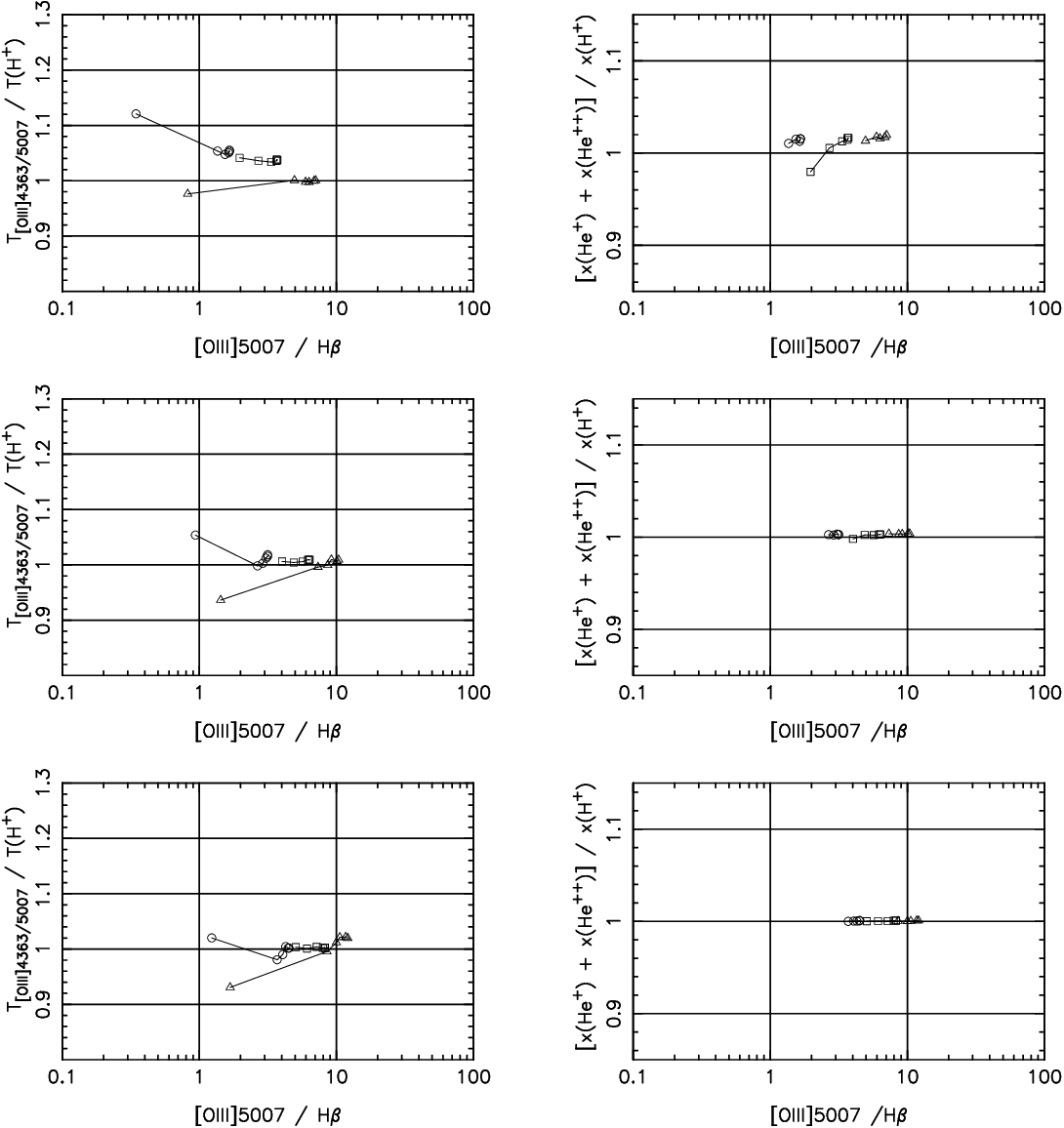,width=15.cm}}
\caption{The same models as in Fig. 4. The symbols have the same meaning as in Fig. 1.
 As in 
Figs. 1 and 4, models shown in the top panels have \mstar = 10$^3$ \Ms, 
those in the middle panels have \mstar = 10$^6$ \Ms, 
those in the bottom panels have \mstar = 10$^9$ \Ms. The left panels concern the 
temperature structure of the models and show the ratio of 
\Toiii/$T$(\Hp) versus \Oiii/\Hb. 
 The right panels concern the ionization structure 
of hydrogen and helium and show ($x$(He$^+$) + $x$(H$^+$))/$x$(H$^+$) 
versus \Oiii/\Hb.
\label{Fig5}}
\end{figure*}
%**************************************

%**************************************
%       Fig6
%**************************************
\begin{figure*}%[htb]
\centerline{\psfig{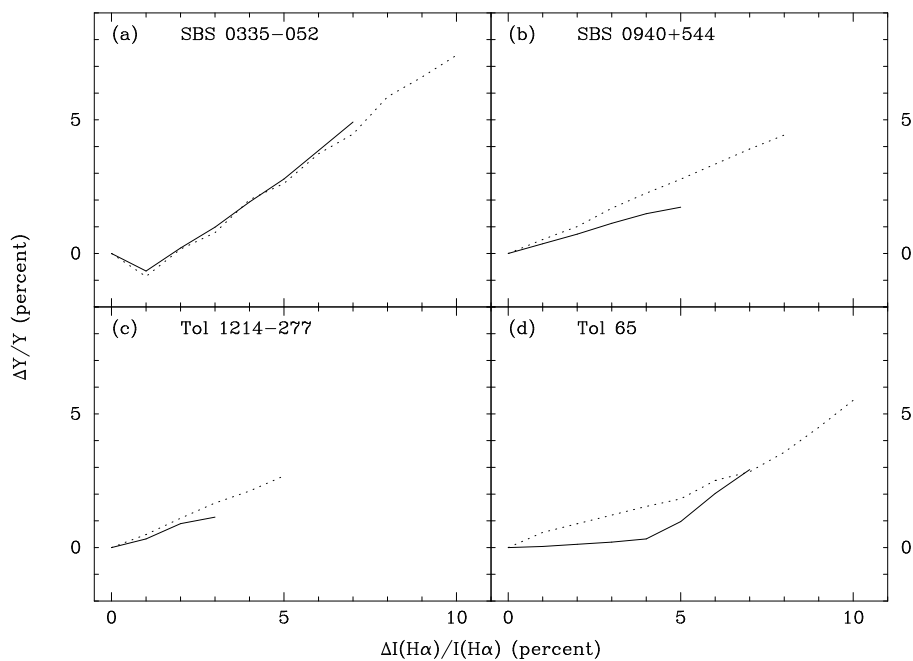}}
\caption{The dependence of the relative correction of the He mass fraction 
$\Delta$$Y$/$Y$ on the contribution of the collisional enhancement 
$\Delta$$I$(H$\alpha$)/$I$(H$\alpha$) of the H$\alpha$ emission line in 
four low-metallicity 
blue compact galaxies. The solid line is the dependence when only the 
intensity of H$\alpha$ emission line is decreased by the amount indicated by 
the abscissa. The dotted line shows the dependence when the H$\alpha$ 
line intensity is decreased by this amount and the H$\beta$ line 
intensity by one third of this amount.
\label{Fig6}}
\end{figure*}
%**************************************

We have selected 4 low-metallicity blue compact galaxies SBS 0335--052 
($Z_\odot$/40, Izotov et al. 2001b), SBS 0940+544 ($Z_\odot$/27, 
Guseva et al. 2001), Tol 1214--277 and Tol 65 ($Z_\odot$/23, Izotov et al. 
2001a) from the sample presented in
Section 3, on the basis of their high-quality spectra obtained
with the Keck telescope and high electron temperatures in the range of 17000 --
20000 K derived from the \oiii\ line intensity ratios. 
Because of the high electron temperature the effect of collisional
excitation on the helium abundance determination is expected to be relatively
large in these galaxies. To derive the $^4$He and heavy element abundances we 
follow the prescriptions by Izotov et al. (1994),
but assuming that there is a certain percentage of collisional excitation in 
the \Ha\ and in the \Hb\ lines. 
We assume that the temperature characteristic 
of the emission of all He {\sc i} lines is given by \Toiii,
and that the ionization correction factor for helium is
equal to 1. This is not exactly true for all our models, as seen
in Fig. 5, which shows (\Toiii/$T$(\Hp) (left panels) and ($x$(\Hep) +
$x$(\Hepp))/$x$(\Hp) (right panels) for the same models as Fig. 4. 
$T$(\Hp) is defined as
\begin{equation}
T_e({\rm H}^+)=\frac{\int{T_en({\rm H}^+)n_edV}}{\int{n({\rm H}^+)n_edV}}
\end{equation}
and $x$(\Hep) is defined as
\begin{equation}
x({\rm He}^{+})=\frac{\int{n({\rm He}^{+})n_edV}}{\int{n({\rm He})n_edV}}.
\end{equation}
 However, since the four considered galaxies have \Toiii\ around 20000 K 
and \Oiii/\Hb\ larger than 3, our model grid indicates that our 
approximations should be reasonable.

In order to explore the effects of collisional excitation of the H lines, we
consider two different cases. In the first case the effect of collisional
excitation is accounted for by decreasing by some amount the intensity of the
H$\alpha$ emission line and leaving unchanged the intensity of the H$\beta$
emission line. In the second case we decrease the intensities of H$\alpha$
and H$\beta$ emission lines assuming that the effect of the collisional
excitation on the H$\beta$ line intensity is three times smaller than that
on the H$\alpha$ line intensity. 
For each assumed value of the contribution of the collisional enhancement to \Ha,
$\Delta$$I$(H$\alpha$)/$I$(H$\alpha$),
we compute the extinction, correct the emission line ratios for reddening 
and underlying stellar absorption, then recompute \Toiii\ 
and derive the $^4$He and the
heavy element abundances. In the
analysis, we use the five strongest He {\sc i} emission lines to derive 
the helium mass fraction $Y$, after correction for collisional and 
fluorescent excitation of the He {\sc i} emission lines. In short, using 
the intensities of the five He {\sc i} emission lines, we evaluate the
electron number density $n_e$ in the He$^+$ zone and the optical depth 
$\tau$($\lambda$3889) in the He {\sc i} $\lambda$3889 line in a self-consistent
way, so that the He {\sc i} $\lambda$3889/$\lambda$5876, 
$\lambda$4471/$\lambda$5876, $\lambda$6678/$\lambda$5876 and
$\lambda$7065/$\lambda$5876 line ratios have their recombination values, after
correction for collisional and fluorescent enhancement of the He {\sc i} 
emission lines. These effects are very strong in hot low-metallicity
H {\sc ii} regions. In particular, the correction of the He {\sc i} 
$\lambda$5876 emission line for collisional and fluorescent enhancement
in SBS 0335--052 is $\sim$ 12\%.

Of course, for each object, there is a maximum 
 contribution of collisional enhancement 
$\Delta$$I$(H$\alpha$)/$I$(H$\alpha$) imposed by the observed 
\Ha/\Hb.  
In reality, the range of possible values of
$\Delta$$I$(H$\alpha$)/$I$(H$\alpha$) is 
smaller because interstellar extinction must be present at some level.
We have checked that, after the completion of the whole procedure,
 the higher order Balmer lines remain compatible with case B within observational errors
and uncertainties in the extinction law.

In Fig. 6 we show the variations of the correction
of the helium mass fraction $\Delta$$Y$/$Y$ derived 
from the He {\sc i} $\lambda$5876 emission line for the four galaxies we
investigated.
It is seen that the correction for the collisional enhancement of the 
hydrogen lines results in an increase of the He abundance in all galaxies. 
Qualitatively such a dependence can be explained by two effects: 1) the 
correction of the H$\beta$ line in the second case (dotted line) increases the 
He {\sc i}/H$\beta$ intensity ratios; 2) the correction of H$\alpha$ and
H$\beta$ lines for collisional excitation decreases the extinction and hence 
increases the intensity
of He {\sc i} $\lambda$5876 emission line. Certainly, some part
of the deviation of the observed H$\alpha$/H$\beta$ intensity ratio 
from the case B value is due to interstellar extinction. Therefore, although
the effect of collisional excitation of the hydrogen lines on the
helium abundance is significant, it is unlikely that the correction 
$\Delta$$Y$/$Y$ exceeds 5\%. At larger values of $\Delta$$Y$/$Y$ the
interstellar extinction derived from the H$\alpha$/H$\beta$ becomes 
very small, less than the foreground extinction produced in our Galaxy.
Our estimate of $\Delta$$Y$/$Y$ is not very certain and should be considered
as a qualitative indication of the relatively large positive correction
of the helium abundance if collisional excitation of the hydrogen 
emission lines is taken into account. 
An accurate determination of the helium abundance 
requires detailed photoionization modeling of the 
H {\sc ii} regions in individual galaxies, with all 
relevant effects taken into account, and with all the 
relevant line intensities satisfactorily reproduced. 
Such an enterprise is still ahead of us.

\section{Conclusion and prospects}

Our consideration of the collisional excitation of hydrogen emission lines
has shown that this effect can be particularly large, 
as high as 8\% for the \Ha/\Hb\ ratio 
in the low-metallicity extragalactic H {\sc ii} regions. If taken into
account it results in an upward correction of the helium abundance $Y$ by
up to  5\%, making the collisional excitation of hydrogen one of the most
important sources of systematics in the primordial helium abundance 
determination. Some other sources of systematic uncertainties
should be considered as well. One of the most important sources may be the
``temperature fluctuations'' which , if present, result in the downward
correction of $Y$ by $\la$ 5\% (e.g., Peimbert et al. 2000). Sauer \&
Jedamzik (2001) considered the effect of the ionization structure and 
large-scale temperature variations in the H {\sc ii} region on the $^4$He
abundance determination. Izotov et al. (2001a) compared the observed
properties of the high-excitation H {\sc ii} regions in the galaxies
Tol 1214--277 and Tol 65 with Sauer \& Jedamzik models and found that a
small downward $Y$ correction (by less than 1\%) should be applied due to the two
mentioned effects. Finally, underlying stellar He {\sc i} line absorption of 
the ionizing clusters should be taken into account. However, this effect is
not large in the four considered blue compact galaxies because of the
large equivalent widths of emission lines and results in an upward $Y$
correction by $\la$ 1\%. 
Combining all these systematic effects  it appears that the helium 
abundance derived by Izotov \& Thuan (1998) and Izotov et al. (1999) in 
low-metallicity blue compact galaxies could be underestimated by about 2 -- 3\%
only.
However, additional work has to be done to clarify the
importance of the systematic effects in the $^4$He abundance determination.
Especially, the mystery of the ``temperature  fluctuations'' has to be solved. 
For this, a detailed modeling of the H {\sc ii} regions in each of the best
observed low-metallicity blue compact galaxies should be done.

%\appendix{}

\begin{acknowledgements}
Y.I.I. has been partly supported by INTAS 97-0033 grant, and acknowledges
support from the ``Kiev project'' of the Universite Paul Sabatier of Toulouse. 
He is grateful for the hospitality of the Paris-Meudon Observatory and of the
Midi-Pyrenees Observatory where this work was conducted.
\end{acknowledgements}

\end{document}